\documentclass{aa}  
\usepackage{graphicx}
\usepackage{txfonts}
\usepackage{hyperref}
\usepackage{caption}
\usepackage{subcaption}
\usepackage[dvipsnames]{xcolor}

\begin{document}

   \title{ALMA reveals a dust-obscured galaxy merger at Cosmic Noon}


   \author{I. Langan
          \inst{1,2},
          G. Popping
          \inst{1},
          M. Ginolfi
          \inst{3,4},
          F. Gentile
          \inst{5,6},
          F. Valentino
          \inst{1,7},
          M. Kaasinen
          \inst{1}
          }

   \institute{European Southern Observatory, Karl-Schwarzschild-Str. 2, D-85748, Garching, Germany\\
              \email{ivanna.langan@eso.org}
         \and
             Univ Lyon, Univ Lyon1, Ens de Lyon, CNRS, Centre de Recherche Astrophysique de Lyon (CRAL) UMR5574, F-69230 Saint-Genis-Laval, France
         \and
             Dipartimento di Fisica e Astronomia, Universit\`a di Firenze, Via G. Sansone 1, 50019, Sesto Fiorentino (Firenze), Italy
         \and
             INAF - Osservatorio Astrofisico di Arcetri, Largo E. Fermi 5, I-50125, Firenze, Italy
         \and
             University of Bologna - Department of Physics and Astronomy "Augusto Righi" (DIFA), Via Gobetti 93/2, I-40129 Bologna, Italy
         \and
             INAF - Osservatorio Astrofisico e Scienza della Spazio, Via Gobetti 93/3, I-40129, Bologna, Italy
         \and
             Cosmic Dawn Center (DAWN)
             }

   \date{Received September 15, 1996; accepted March 16, 1997}

 
  \abstract
  {Galaxy mergers play a critical role in galaxy evolution - altering the size, morphology, dynamics and composition of galaxies. So far, galaxy mergers have mostly been identified through visual inspection of their rest-frame optical and NIR emission. But, dust can obscure this emission, resulting in the misclassification of mergers as single galaxies, and the incorrect interpretation of their baryonic properties.}   
   {Having serendipitously discovered a dust-obscured galaxy merger at $z=1.17$, we aim to determine the baryonic properties of the two merging galaxies, including the star formation rate, and stellar, molecular gas, and dust masses.}
  {Using Band 3 and 6 observations from the Atacama Large Millimeter and submillimeter Array (ALMA), and ancillary data, we study the morphology of this previously misclassified merger. We deblend the emission, derive the gas masses from CO observations, and model the spectral energy distributions, to determine the properties of each galaxy. Using the rare combination of ALMA CO(2--1), CO(5--4) and dust-continuum (rest-frame 520$\mu$m) observations, we provide insights into the gas and dust content and ISM properties of each merger component. }
  {We find that only one of the two galaxies is highly dust-obscured, whereas both are massive ($>10^{10.5} M_\odot$), highly star-forming (SFR$=60-900 M_\odot$/yr), have a moderate-to-low depletion time ($t_\text{depl}<0.7$Gyr) and high gas fraction ($f_\text{gas}\geq1$).}
   {These properties can be interpreted as the positive impact of the merger. With this serendipitous discovery, we highlight the power of (sub)millimeter observations to identify and characterise the individual components of obscured galaxy mergers.}

   \titlerunning{dust-obscured galaxy merger}
   \maketitle
%

\section{Introduction}
 Throughout their evolution, galaxies can encounter one or more companions close enough that the gravitational interaction pulls them together in \textit{galaxy mergers}. During galaxy mergers, the evolution of galaxies is driven by turbulent and stochastic processes. This can impact dramatically the properties of galaxies on many levels including star formation (e.g., \citealt{Ellison2022}), AGN activity (e.g., \citealt{ByrneMamahit2023}) and morphology (e.g., \citealt{Martin2018}). In particular, mergers can entrain gas to the center of galaxies by breaking the angular momemtum of the accreting gas, causing for instance, elevated star formation with respect to typically star-forming galaxies along the main-sequence (e.g., \citealt{Kim2009}, \citealt{Saitoh2009}, \citealt{Kaviraj2015}, \citealt{Tacchella2016}, \citealt{Pearson2019}).\par

Merger rates appear to increase as a function of redshift (e.g., \citealt{Ventou2019}, \citealt{Romano2021}, \citealt{Conselice2022}, \citealt{Ren2023}). In the local Universe, close to 1\% of galaxies with similar masses (i.e., major mergers) are merging, while the fraction goes up to just below 20\% of galaxies  at Cosmic Noon, i.e., $z = 1-2$, where the Universe is the most active and reaches its peak of star formation (e.g., \citealt{Madau2014} and references therein). Based on a visual classification from rest-frame V-band Hubble Space Telescope (HST) imaging, \citet{Kaviraj2013} found that major mergers contribute up to 27\% to the star formation activity at the start of Cosmic Noon, i.e., $z\sim2$. Furthermore, galaxy mergers are a unique laboratory to study the different processes involved in the evolution of galaxies, because they bridge a wide range of physical scales from star formation (e.g., enhancement of star formation due to gas and dust compression at sub-pc scales) to large-scale structures (galaxy clustering at Mpc scales).\par

Galaxy mergers are key to understand the evolution of galaxies. However, accurately identifying them all remains a challenge. In works based on observations (e.g., \citealt{Mundy2017}, \citealt{Ren2023}), the common method to identify galaxies as mergers is to visually inspect rest-frame optical or NIR data using telescopes such as HST, with a set of selection criteria based on spatial and velocity separations to ensure that the galaxies are gravitationally bound. Typically galaxies are selected, from rest-frame optical or NIR data, to lie within a few tens of kpc of each other or with a relative velocity of less than a few hundreds of km s$^{-1}$ (e.g., \citealt{Lotz2008}, \citealt{Casteels2014}, \citealt{Ventou2019}).
The common method described above implies that we are able to identify \textit{all} the galaxies part of merging systems to classify systems as such, using rest-frame optical or NIR data alone. However, at $z\sim1-2$, nearly $70\%$ of the ongoing star formation is in an obscured phase (\citealt{Zavala2021}). This can result in obscured galaxies being missed if only rest-optical observations are used. For instance, \citet{Talia2021}, \citet{Enia2022}, \citet{Behiri2023}, \citet{Smail2023}, \citet{Gentile2023} found dusty star-forming galaxies (DSFGs) that appeared optically dark because of dust obscuring the emission at rest-frame optical wavelengths. Galaxy mergers could thus be missed by classical merger identification approaches, because one or several members of the merger can originally (i.e., before the mergering phase) be dusty or because of the merger itself driving up the build up of a large dust reservoir.
If such optically-dark mergers exist, we could observe them at longer wavelength,  i.e., the mid-infrared to (sub)millimetre wavelengths, where we are not limited by obscuring dust. With the advance of telescopes capable of observing at these longer wavelengths and reaching high angular resolution, such as the \textit{Atacama Large Millimeter and submillimeter Array} (ALMA), we can thus reveal very complex systems that were otherwise hidden by dust. These systems represent unique laboratories to further our understanding of the baryonic properties of galaxies. In particular, CO observations with sufficient angular resolution to disentangle merger members provide us with a window into the molecular gas and dust content of individual galaxies in mergers at Cosmic Noon. Thus, (sub)millimeter observations improve the characterisation of mergers by adding information otherwise unaccessible at shorter wavelengths (e.g., NIR).\par

Using ALMA archival observations, we present the serendipitous detection of a system within the Cosmic Evolution Survey (COSMOS, \citealt{Scoville2007}) field consisting of two massive merging galaxies at $z\sim1$. This system was previously misclassified as a single source (COSMOS-51599) because one of the two galaxies is optically dark, that is, no emission is apparent in the HST WFC3/UVIS F814W observations (COSMOS "super-deblended" catalogue, \citealt{Jin2018}), COSMOS2020 catalogue, \citealt{Weaver2022}). In this work we use ALMA archival observations to reveal an additional galaxy, indicating the presence of a previously-missed merger system (which we refer to as Matilda\footnote{The nickname comes from the name of the cat living at the base camp of the Atacama Pathfinder Experiment telescope where the work presented in this paper started.}). We use the archival ALMA CO(2--1), CO(5--4) and dust continuum observations to study the gas and dust content and ISM physical conditions of the two galaxies involved in the merger.\par

This paper is organised as follows. In Section \ref{section:observations}, we present the observations and data reduction. In Section \ref{section:analysis}, we describe how we identified the system and how we derived the baryonic properties of each galaxy. In Section \ref{section:discussion}, we discuss the results and their implications. We summarise our findings in Section~\ref{subsec:summary}. The cosmology assumed throughout this work follows the $\Lambda$CDM standard cosmological parameters: $H_0 = 70 \text{km}\, \text{s}^{-1}\, \text{Mpc}^{-1}$, $\Omega_m = 0.3$ and $\Omega_\Lambda = 0.7$ \citep{Planck2020}. We use a \citet{Chabrier2003} stellar Initial Mass Function (IMF).

\section{Multi-wavelength observations and data reduction}
\label{section:observations}

\subsection{ALMA}
\label{subsec:ALMAdata}

We use the calibrated raw visibility data from the observations of programmes 2015.1.00260.S and 2016.1.00171.S (PI: Daddi), provided by the European ALMA Regional Centre \citep{Hatziminaoglou2015}. These two programmes include observations of Matilda (COSMOS-51599), at RA 09:58:23.630 and DEC +02:12:01.660, in Band 3 ($2.6-3.6$mm) and Band 6 ($1.1-1.4$mm), corresponding to CO(2--1), CO(5--4) and dust-continuum. Matilda is part of a sample of galaxies presented in \citet{Valentino2020} and we refer the reader to this work for more details on the design of the survey and the associated observations.\par

\begin{figure*}
\sidecaption
    \includegraphics[width=12cm]{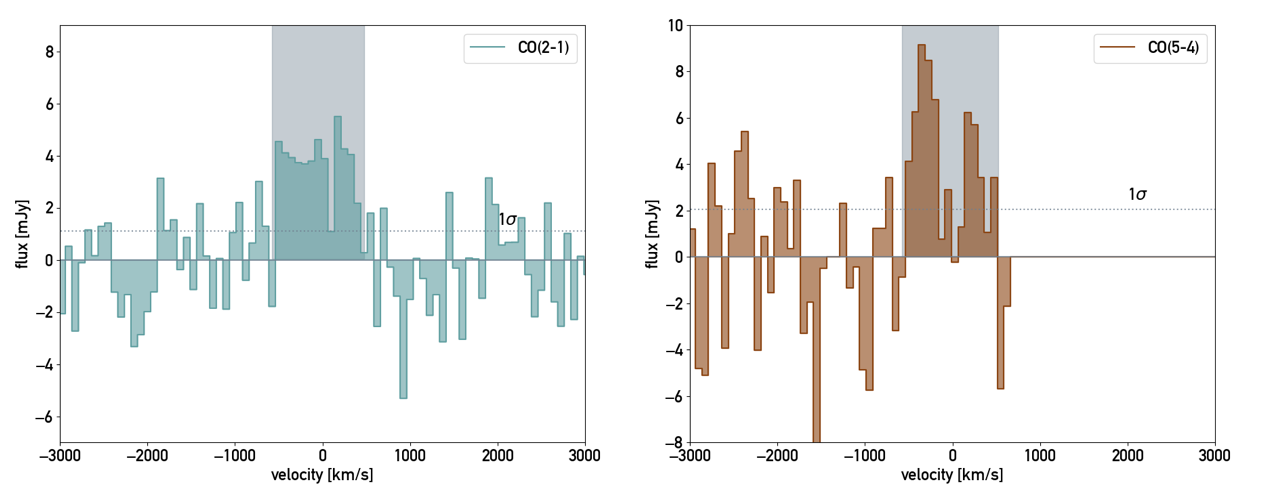}
        \caption{The CO(2--1) (left) and CO(5--4) (right) emissions of the entire system in mJy per 75 km/s as a function of the velocity, where the center velocity is determined from the redshift $z=1.17$. The channels used to create the intensity maps are highlighted with the grey shaded area.}
        \label{fig:spectra}
\end{figure*}
\begin{figure*}
\centering
    \includegraphics[width=\textwidth]{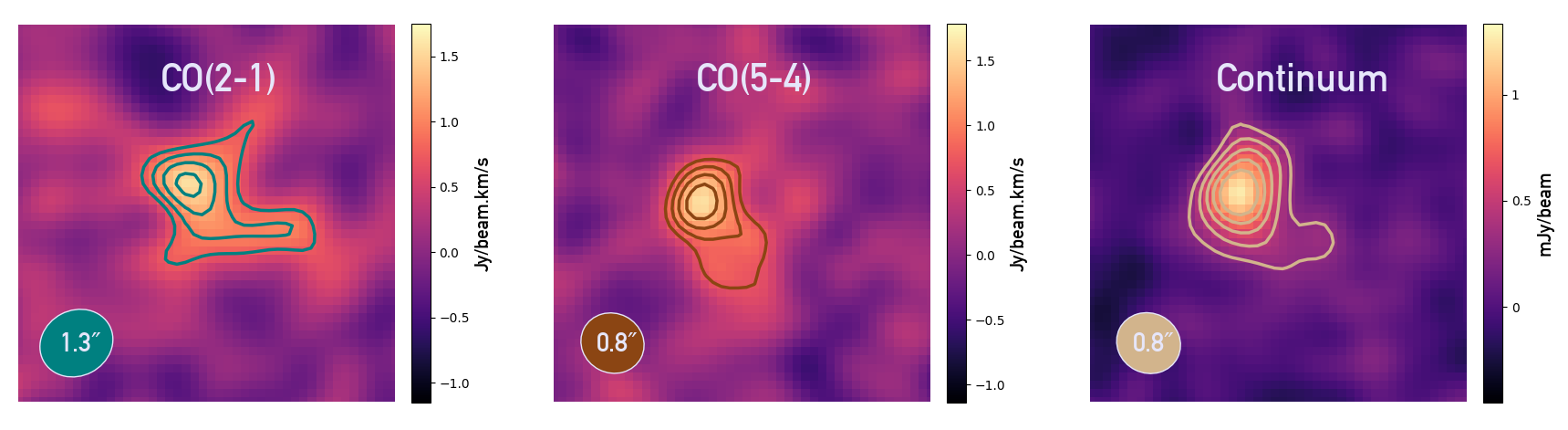}
        \caption{Intensity maps ($7\arcsec \times 7\arcsec$) of the CO(2--1) emission (left), with [3, 4, 5, 6] $\sigma_{\mathrm{CO(2-1)}}$ contours, the CO(5--4) emission (middle), with [3, 4, 5, 6] $\sigma_{\mathrm{CO(5-4)}}$ contours, and the dust-continuum image (right), with [3, 5, 7, 9, 11] $\sigma_{\mathrm{dust}}$ contours.}
        \label{fig:Moment0maps}%
\end{figure*}

We use the Common Astronomy Software Applications (CASA) data processing software \citep{CASA2022} throughout the analysis of the ALMA data. For the CO(2--1) observations, we perform uv-plane continuum subtraction (with \texttt{uvcontsub}), and image the CO(2--1) emission with natural weighting and channel width $\Delta v = 75$ km s$^{-1}$ (with \texttt{tclean}). This imaging step results in a cleaned cube, with a sensitivity of $0.82$ mJy/beam in channels of 75 km s$^{-1}$, where the beam is $1.5\arcsec \times 1.3\arcsec$. We also create the intensity map of the CO(2-1) emission (with \texttt{immoments}) by integrating between -575km/s to +475km/s (channels 55 to 69), where the line clearly shows in a spectrum extracted from an aperture of $2\arcsec$ encompassing the entire system (see left panel of Figure \ref{fig:spectra}).
For the CO(5--4) data, we repeat the same procedure, that is we subtract the continuum and image with the same set of parameters as for the CO(2--1) data. This yields a CO(5--4) cube with a sensitivity of $0.81$ mJy/beam in channels of 75 km s$^{-1}$, where the beam is $0.8\arcsec \times 0.7\arcsec$. To create the intensity map of the CO(5--4) emission, we integrate this emission from -605km/s to +445km/s (channels 38 to 52). We choose this range because this is where the line clearly shows in a spectrum extracted with the same $2\arcsec$ aperture as for the CO(2--1) emission (see right panel of Figure \ref{fig:spectra}).
We image the continuum from the Band 6 data, excluding the region that covers the CO(5--4) line, resulting in a continuum image with a sensitivity of $0.23$ mJy/beam where the beam is $0.8\arcsec \times 0.7\arcsec$. All the intensity maps are shown in Figure \ref{fig:Moment0maps}. The CO(2--1), CO(5--4) and continuum emissions are clearly detected with a peak-SNR of 6, 6 and 14, respectively.\par

%

\subsection{Other observations}


Because our target is in the COSMOS field (\citealt{Scoville2007}), multi-wavelength ancillary data is already available. In this work, we make use of the ancillary photometric images available for our target through the IRSA COSMOS cutout service\footnote{\url{https://irsa.ipac.caltech.edu/data/COSMOS/index_cutouts.html}}, including \textit{Subaru}/HSC data in the g, r, z and y filters (\citealt{Aihara2019}), VISTA data in the Y, J, H and Ks filters (\citealt{McCracken2012}), HST/WFC in the F814W filter (\citealt{Koekemoer2007}, \citealt{Massey2010}), \textit{Spitzer}/IRAC in all channels (\citealt{Moneti2021}), and \textit{VLA} at 3 GHz (10 cm) observations (\citealt{Smolcic2017}). We note that \citet{Weaver2022} astrometrically corrected all datasets based on Gaia DR2 (\citealt{Gaia2018}), i.e., the HST data is aligned with the other data sets. Similarly, the ALMA observations are astrometrically aligned with the other datasets with an offset of up to 23 mas (section 10.5.2 of \cite{almahandbook}). This offset is negligible compared to the beam size of the observations used in this work (0.8”, in the case of the CO(5-4) observations and 2” in the case of the CO(2-1) observations).



\section{Analysis of the data}
\label{section:analysis}

\subsection{Identification of the dusty galaxy merger}

\begin{figure}
    \resizebox{\hsize}{!}{\includegraphics{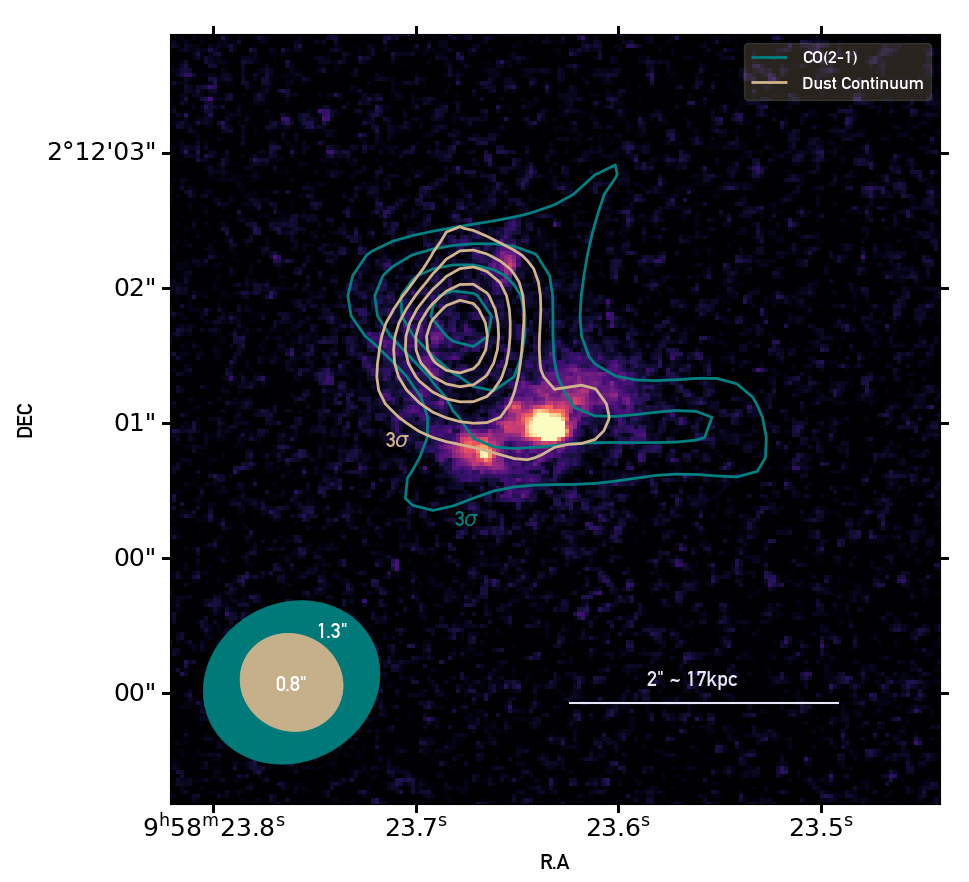}}
    \caption{HST F814W image from \citet{Koekemoer2007}, with ALMA CO(2--1) contours in blue at [3, 4, 5, 6] $\sigma_{\mathrm{CO(2-1)}}$ and ALMA dust-continuum in beige at [3, 5, 7, 9, 11] $\sigma_{\mathrm{dust}}$. The beams are shown in the lower left, with sizes of 1.3$\arcsec$ and 0.8$\arcsec$ for the CO(2--1) and dust-continuum emissions, respectively. A scale is shown at the bottom right to represent the physical scales at the redshift of the system, i.e., $z=1.17$.}
    \label{fig:almaoverhst}
\end{figure}

\begin{figure*}
\centering
    \includegraphics[width=\textwidth]{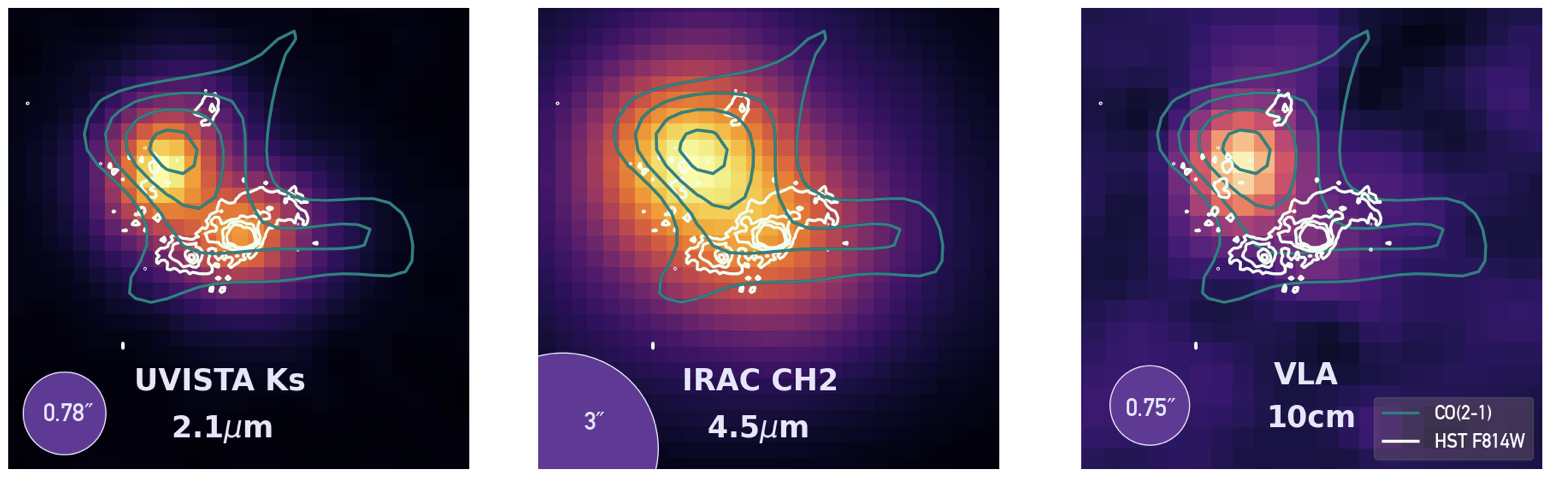}
        \caption{ALMA CO(2--1) emission contours (in blue) and HST/F814W emission contours (in white) overlaid on Ultravista Ks band (left), IRAC channel 2 (middle), and VLA 10 cm (right) observations. These images are post stamps of $5\arcsec \times 5\arcsec$.}
        \label{fig:almahstoverothers}
\end{figure*}

As shown in Figure \ref{fig:Moment0maps}, the CO(2--1) emission contours show an irregular morphology (i.e., inconsistent with a simple beam shape), with the bulk of the emission concentrated in the 6 $\sigma$ contours in the North and a tail of fainter emission at 3 and 4 $\sigma$ towards the South. The CO(5--4) and dust-continuum contours also show a disturbed morphology (i.e., the emission extends, in one direction, beyond the beam shape), albeit less strongly than the CO(2--1) emission. We compare the extent of the CO(2--1) and dust-continuum emissions to the stellar emission traced by the HST/F814W data in Figure \ref{fig:almaoverhst}. We find that the CO(2-1) emission extends beyond the HST/F814W emission, covering $\sim2.6\arcsec$ (i.e., $\sim20\,$kpc) and peaks where the emission in the HST/814W observations is low. We also find an offset of the emission traced by ALMA with respect to the HST observations. The peak of the ALMA CO(2--1) emission is $\sim1\arcsec$ away from the peak in the HST emission, corresponding to physical scales of 8kpc. 
We compare the extent and peak of the CO to the VISTA/Ks band, Spitzer/IRAC channel 2 and VLA/3GHz observations (see Figure \ref{fig:almahstoverothers}). The VISTA Ks-band observations, which trace old stellar populations, have two peaks of emission, one coincident with the galaxy visible in the HST/F814W data (white contours) and the other with the peak of the ALMA CO(2--1) and CO(5--4) emissions (only the ALMA CO(2--1) emission is shown, with the blue contours, for visual clarity). In the Spitzer/IRAC channel 2 ($4.5\mu m)$ data, which also trace old stars, the PSF is too large to distinguish different components (see also \citealt{Jin2018}, \citealt{Valentino2020}). The VLA/3GHz data, tracing high-energy sources such as AGN and dust-unbiased star formation, is completely offset from the HST data but fully consistent with the peak of the ALMA data, confirming once more the presence of another galaxy missed in previous work.
Thus, we conclude that we are witnessing two galaxies at $z=1.17223 \pm 0.00037$ (\citealt{Valentino2020}) merging. For the remainder of the paper, we call North the northern component (invisible in the HST/F814W data, bright in the VLA/3GHz data and consistent with the peak of the ALMA data) and South the southern component (visible in the HST/F814W data and invisible in the VLA data).


\begin{figure*}
\centering
    \begin{subfigure}[b]{\textwidth}
        \centering
         \includegraphics[width=\textwidth]{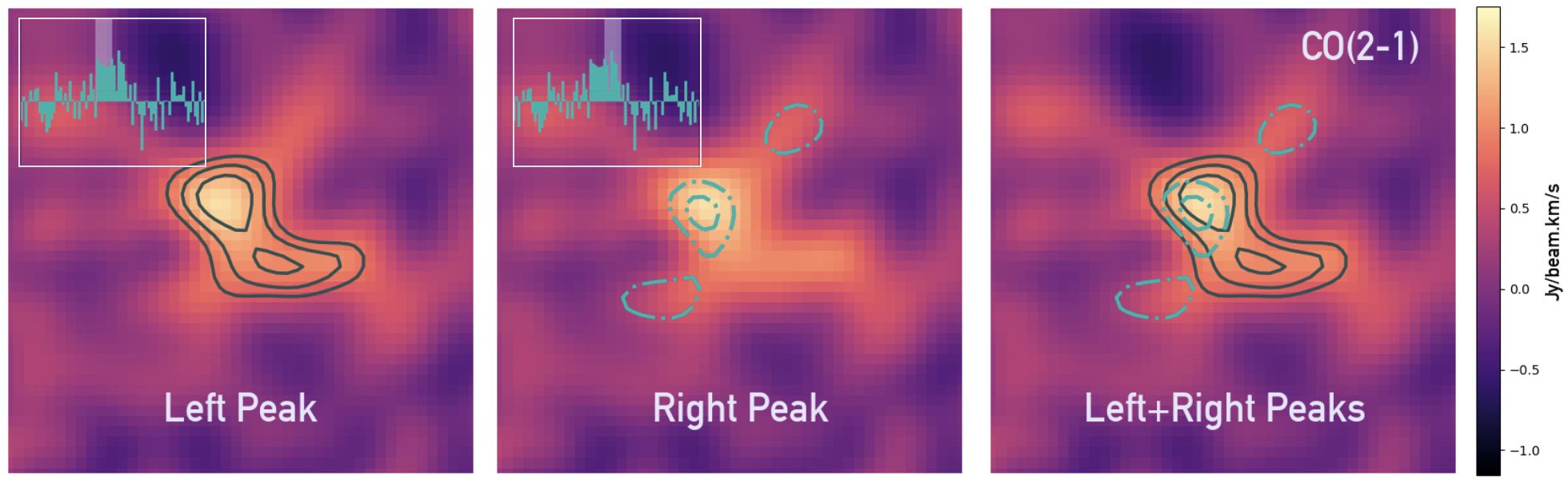}
     \end{subfigure}
     \begin{subfigure}[b]{\textwidth}
        \centering
         \includegraphics[width=\textwidth]{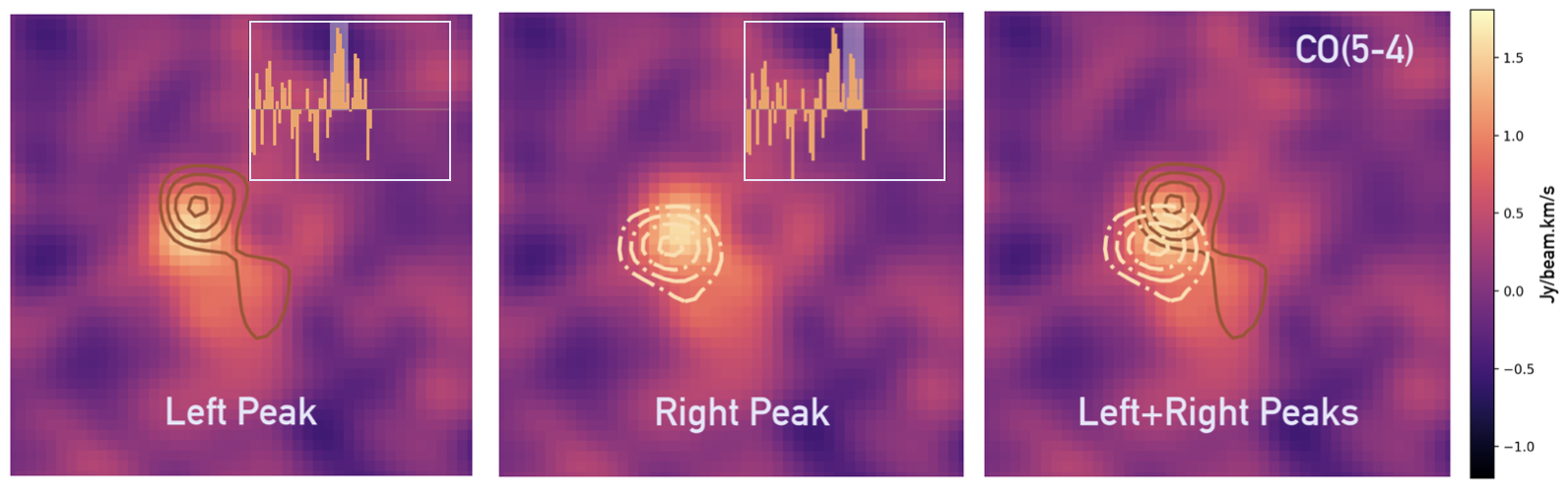}
     \end{subfigure}
\caption{Contours of the CO(2--1) (top row) and CO(5--4) (bottom row) intensity maps, when integrating around the negative velocities peak (left column, solid lines) or the right peak (middle column, dash-dotted lines) of the emissions. The right column shows the two peaks overlaid. The negative and positive velocities peaks correspond to -575km/s and -50km/s and -50km/s to -475km/s, respectively, for both CO(2--1) and CO(5--4) emissions. The contours start at 3 $\sigma$ and increase with an indent of 1 $\sigma$. The background image in the top row (bottom row) is the CO(2--1) (CO(5--4)) emission when integrating the entire peak, as shown in the left (right) panel of Figure \ref{fig:Moment0maps}. An inset shows the spectrum and the channels used in the imaging are highlighted with the grey shaded area. }
\label{fig:PeaksMom0}
\end{figure*}

\subsection{Velocity and spatial offsets of the dusty galaxy merger}
\label{subsec:dynamics}

We image the velocity offset in the CO(2--1) and CO(5--4) emissions shown in Figure \ref{fig:spectra}. We image the negative velocities peak (left column, solid lines in Figure \ref{fig:PeaksMom0}) by integrating between -575km/s and -50km/s and the positive velocities peak (middle row, dash-dotted lines in Figure \ref{fig:PeaksMom0}) by integrating between -50km/s and +475km/s, for both CO(2--1) and CO(5--4) (shown in top row and bottom row of Figure \ref{fig:PeaksMom0}, respectively).

As shown in the right panels of the top and bottom rows of Figure \ref{fig:PeaksMom0}, North is traced by both peaks. In both the CO(2--1) and the CO(5--4) emissions, we see a slight shift between the negative and positive velocities peaks at the location of North. On the other hand, South seems to only be traced by the negative velocities peak, both in the CO(2--1) and CO(5--4) emissions. Despite observing two clear peaks in the CO(5--4) spectrum, when imaging them separately as shown in the bottom row of Figure \ref{fig:PeaksMom0}, we do not observe a clear correspondence between one peak and its location in the imaging, which could be expected if one peak represented one galaxy. The same applies for CO(2--1), albeit a less obvious distinction between the two peaks in the spectrum.

\subsection{The CO flux measurements}
\label{subsec:coflux}

To disentangle the CO emission of the two galaxies part of the merger, we adopt the following deblending approach. 
We compute the CO(2--1) flux of the entire system within the 3 $\sigma$ contours shown in the left panel of Figure \ref{fig:Moment0maps}. From that CO(2--1) intensity map, we see bright emission within the size of a one-beam element centred on North. Therefore, we associate this bright component to North and estimate the CO(2--1) flux by measuring the flux within the single-beam element centred on North. The corresponding flux error is the standard deviation of the CO(2--1) intensity map, masking the emission of the merger (i.e., the emission within the $2\arcsec$ used to extract the spectrum of the merger in Section \ref{subsec:ALMAdata}). This resulted in $F_{\mathrm{CO(2-1)}}^\mathrm{North} = 1.56 \pm 0.25$ Jy km/s. We associate the remaining emission within the CO(2--1) 3 $\sigma$ contours to South. That remaining emission being too faint, we used a 3 $\sigma$ upper limit, resulting in $F_{\mathrm{CO(2-1)}}^\mathrm{South} \leq 0.75$ Jy km/s.
With this method, we make the assumption that all the gas that is not associated with North belongs to South. There could be gas in between the two interacting galaxies making this assumption incorrect. But given the limited resolution of the observations, we are forced to make such simplifying assumptions. We performed this "deblending method" instead of using the PhoEBO (\citealt{Gentile2023}) code used later in Section \ref{subsec:SEDfitting} because this code requires priors based on the position of the stellar emission, which we cannot assume to align with the molecular gas and dust.\par
To derive the CO(5--4) fluxes of the two merging galaxies, we follow the same procedure. We associate the peak of the CO(5--4) emission within the 3 $\sigma$ contours of the CO(2-1) intensity map emission to North, resulting in $F_{\mathrm{CO(5-4)}}^\mathrm{North} = 1.61 \pm 0.24$ Jy km/s, with the error being the standard deviation of the CO(5--4) intensity map (masking the emission coming from the merger within $2\arcsec$). As for the CO(2-1) emission, the remaining CO(5-4) emission within the 3 $\sigma$ CO(2-1) contours is too faint, therefore we used a 3 $\sigma$ upper limit for the Southern component, $F_{\mathrm{CO(5-4)}}^\mathrm{South} \leq 0.72$ Jy km/s. We proceed in the same way for the dust-continuum emission, finding for North $F_{\mathrm{cont}}^\mathrm{North} = 1.26 \pm 0.09$ mJy and for South $F_\mathrm{cont}^\mathrm{South} \leq 0.28$ mJy.

\subsection{The molecular gas masses from the CO(2-1) emission}
\label{subsec:mh2}
We derive the molecular gas masses of the two galaxies by first converting the CO(2--1) fluxes to line luminosities, via

\begin{equation}
L_{\mathrm{CO(2-1)}} = 3.25 \times 10^7 \times F_{\mathrm{CO(2-1)}} \times \frac{D_L^2}{(1+z)^3 \nu_{obs}^2} \mathrm{\, \, \,K \,km \,s^{-1} \,pc^2}
\label{eq:fluxtolum}
\end{equation}
We convert the CO(2-1) luminosities to CO(1--0) luminosities, assuming a ratio of 0.85 between the CO(2--1) and CO(1--0) luminosities (\citealt{Carilli2013}). Then, we derive the molecular gas masses using the molecular gas mass conversion factor $\alpha_{\text{CO}}$, with $\alpha_{\text{CO}} = 3.4 \pm 2$ $\mathrm{(K\,km\,s^{-1}\,p^2)^{-1}}$, the mean and standard deviation measured for a sample of $z=1-3$ DSFGs in \citet{Harrington2021}. The upper and lower errors on the molecular gas masses (see Table \ref{table:properties}) reflects the uncertainty on $\alpha_{\text{CO}}$. We caution that the resulting molecular gas masses are highly dependent on the adopted $\alpha_{\text{CO}}$ value. Lower values, e.g., $\alpha_{\text{CO}} \sim 0.9$ as suggested in \citet{Bolatto2013} for starbursts, would result in 5 times lower molecular gas masses, therefore also impacting the results discussed in Section \ref{subsec:context}.

\subsection{The molecular ISM excitation conditions}

We measure a CO(5--4) to CO(2--1) line ratio of $F_{\mathrm{CO(5-4)}}/F_{\mathrm{CO(2-1)}} = 1.03 \pm 0.35$ for North. The flux measurements for South are upper limits, making $F_{\mathrm{CO(5-4)}}/F_{\mathrm{CO(2-1)}}$ highly uncertain, therefore we omit it for this galaxy. The CO(5--4) to CO(2--1) ratio of North is similar to what \citet{Boogaard2020} find for a stack of 22 SFGs at $<z>=1.2$, that is $F_{\mathrm{CO(5-4)}}/F_{\mathrm{CO(2-1)}} = 1.41 \pm 0.15$. However, works on Submillimeter Galaxies (SMGs) at high redshift ($z>2$) (e.g., \citealt{Bothwell2013}, \citealt{Spilker2014}) find $F_{\mathrm{CO(5-4)}}/F_{\mathrm{CO(2-1)}} > 2.4$. In addition, \citet{Valentino2020} find an increase of the CO(5--4) to CO(2--1) ratio with distance to the main-sequence for a sample of a few tens of galaxies at $<z>=1.25$, with $F_{\mathrm{CO(5-4)}}/F_{\mathrm{CO(2-1)}} = 1.6 \pm 0.2$ for MS galaxies to $F_{\mathrm{CO(5-4)}}/F_{\mathrm{CO(2-1)}} = 2.2 \pm 0.3$ for extreme starburst galaxies. Therefore, North exhibits ISM excitation conditions similar to MS galaxies at similar redshifts.

\subsection{The stellar mass, dust mass and star formation rate from SED fitting}
\label{subsec:SEDfitting}
To measure the stellar mass, dust mass, and star formation rate (SFR), we separately model the spectral energy distributions (SEDs) of the two galaxies identified to be part of the dusty galaxy merger, by fitting the photometry with the SED-modelling tool Multi-wavelength Analysis of Galaxy Physical Properties (\texttt{MAGPHYS}, \citealt{daCunha2015}, \citealt{Battisti2020}). However, we first need to deblend the two galaxies. The deblending is especially required for the Spitzer/IRAC observations where the resolution is insufficient to distinguish the two galaxies.\par

We extract the photometry of Matilda from the maps employed by \citet{Weaver2022} to assemble the COSMOS2020 catalogue. These data include the optical, NIR, and MIR maps from Subaru/HSC, \textit{VISTA}/VIRCAM, and \textit{Spitzer}/IRAC instruments and telescope, respectively. To account for the significant source blending between the two components of this dusty galaxy merger, we use “Photometry Extractor For Blended Objects” (\texttt{PhoEBO}; \citealt{Gentile2023}). This code implements a slightly modified version of the algorithm introduced by \citet{Labbe2006} and already employed in several studies in the current literature (see e.g., \citealt{Endsley2021}, \citealt{Whitler2023}), but optimized for the deblending of the so-called Radio-Selected NIRdark galaxies (i.e. sources with a radio counterpart and no detection at optical/NIR wavelengths, e.g. \citealt{Talia2021}, \citealt{Enia2022}, \citealt{Behiri2023}, \citealt{Gentile2023}). PhoEBO deblends the two galaxies making up Matilda using a double prior coming from the “high-resolution” images in the radio and in the NIR bands, employing a PSF-matching with the “low-resolution” images (mainly those in the four IRAC channels) to attribute the flux to the different components present in the analysed system. A detailed description of the code, available here\footnote{\url{https://github.com/fab-gentile/PhoEBO}}, can be found in \citet{Gentile2023}. To measure the flux of the two deblended components, we perform a standard aperture photometry on the two galaxies with Photutils (\citealt{Bradley2023}), employing a fixed diameter of $4\arcsec$ and locally subtracting the background noise evaluated in a two arcsec-wide annulus around each source. The uncertainties are computed by photutils by adding in quadrature the noise (obtained from the weight maps) for all the pixels in the considered apertures. To allow the SED-fitting code to explore a wider range of properties (see e.g \citealt{Laigle2016}, \citealt{Weaver2022}), and to account for possible systematics in the photometry extractions (see the discussion in \citealt{Gentile2023}), we add in quadrature a constant value of 0.15 mag to the uncertainties reported by PhoEBO before giving them in input to the SED-fitting code. An example of the output deblended maps of PhoEBO is shown in the appendix \ref{section:apd}. We include the ALMA band 6 dust-continuum data in the SED fits, which we obtained as described in Section \ref{subsec:coflux}. The results of the photometry are given in Table \ref{table:fluxesfromPHOEBO}.\par


\begin{table}
\caption{Fluxes resulting from PhoEBO and our ALMA analysis.}       
\label{table:fluxesfromPHOEBO}      
\centering  
\renewcommand{\arraystretch}{1.5} 
\begin{tabular}{c c c c}      
\hline\hline                 
Instrument & North & South\\
\hline                        
   HSC/g & $0.45 \pm 0.15$ & $1.31 \pm 0.15$\\
   HSC/r & $0.79 \pm 0.16$ & $2.26 \pm 0.16$\\
   HSC/i & $1.57 \pm 0.16$ & $4.17 \pm 0.16$\\
   HSC/z & $3.11 \pm 0.17$ & $7.78 \pm 0.17$\\
   HSC/Y & $3.94 \pm 0.21$ & $9.83 \pm 0.21$\\
   VISTA/Y & $3.68 \pm 0.42$ & $7.55 \pm 0.42$\\
   VISTA/J & $7.02 \pm 0.47$ & $10.83 \pm 0.47$\\
   VISTA/H & $14.00 \pm 0.61$ & $15.92 \pm 0.61$\\
   VISTA/Ks & $31.44 \pm 0.43$ & $25.79 \pm 0.43$\\
   IRAC/Ch1 & $86.30 \pm 0.20$ & $34.88 \pm 0.20$\\
   IRAC/Ch2 & $91.44 \pm 0.24$ & $30.20 \pm 0.25$\\
   IRAC/Ch3 & $77.38 \pm 3.34$ & $17.13 \pm 0.35$\\
   IRAC/Ch4 & $68.10 \pm 4.83$ & $24.54 \pm 4.86$\\
   ALMA/B6 & $1260 \pm 90$ & $\leq280$*\\
\hline                                   
\end{tabular}
\tablefoot{Fluxes and errors in $\mu$Jy , for North (second column) and South (third column), given as inputs in the SED fits discussed in Section \ref{subsec:SEDfitting}. * indicates a 3 $\sigma$ upper limit. The first column is the instrument and its band (instrument/band).}
\end{table}

\begin{figure}
    \centering
    \begin{subfigure}[b]{\columnwidth}
         \centering
         \includegraphics[width=\columnwidth]{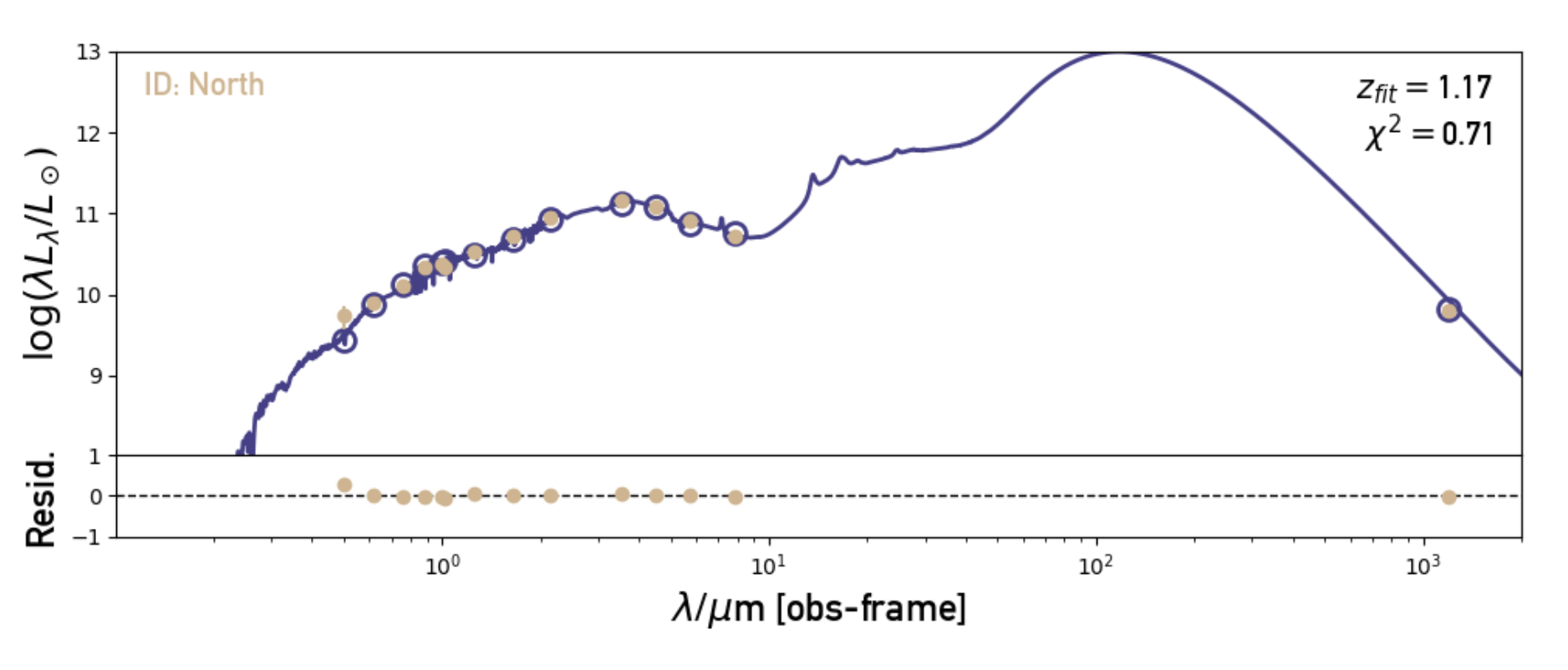}
     \end{subfigure}
     \begin{subfigure}[b]{\columnwidth}
         \centering
         \includegraphics[width=\columnwidth]{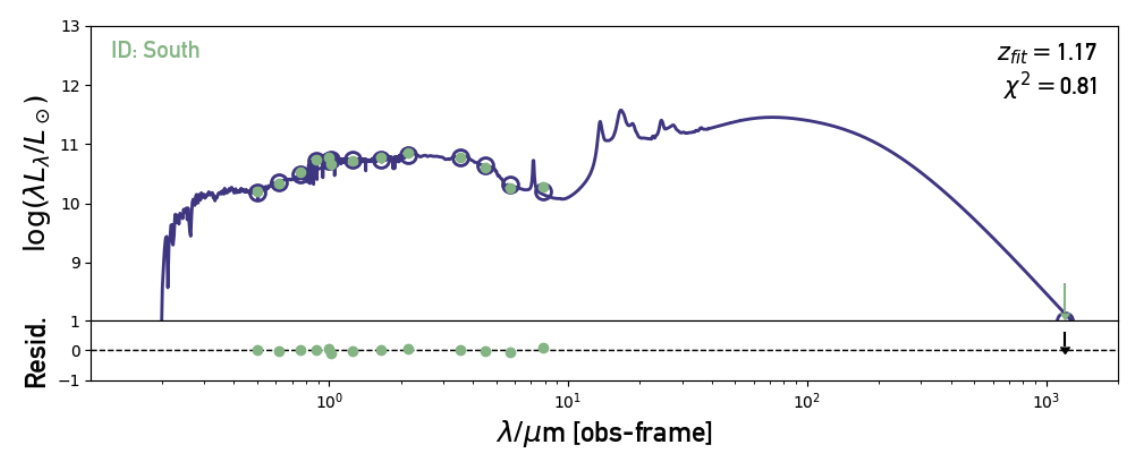}
     \end{subfigure}
     \caption{SED fitting results for North (top) and South (bottom). The model fitted is shown with the purple line and the observations are shown with the beige (North) or green (South) dot markers. For South, the ALMA upper limit observation at $1.2\mu m$ is shown with an arrow pointing down.}
     \label{fig:SED}
\end{figure}

From these photometric results, we perform SED fitting with \texttt{MAGPHYS}, fixing the redshift to $z=1.17223$.
In Figure \ref{fig:SED} we show the results from the SED fitting analysis for the best-fit models with $\chi^2 = 0.71$ and $\chi^2 = 0.81$ for North and South, respectively. We summarise the output of the SED fits in Table \ref{table:properties}. We find that North has a stellar mass $\log(M_*/M_\sun)_N = 10.95^{+0.13}_{-0.05}$,  a dust mass $\log(M_{\text{dust}}/M_\sun)_N = 8.82^{+0.09}_{-0.07}$ and a star formation rate $\text{SFR}_N = 933^{+90}_{-120} M_\sun$/yr. South has $\log(M_*/M_\sun)_S = 10.59^{+0.08}_{-0.08}$, $\log(M_{\text{dust}}/M_\sun)_S = 7.56^{+0.30}_{-0.33}$ and $\text{SFR}_S = 60^{+42}_{-29} M_\sun$/yr. As a consistency check, we calculated the SFR from the VLA/3GHz observations available (following equation 12 from \citealt{Kennicutt2012}). We found $\text{SFR}_{N, VLA} = 376.8 \pm 6.3 M_\sun$/yr for North based on the single source detection reported in \cite{Smolcic2017} and $\text{SFR}_{S, VLA} \leq 46 M_\sun$/yr based on the 3 $\sigma$ noise limit of the VLA 3GHz map for South. The results between the SFRs resulting from SED fitting and the ones based on the VLA data are thus consistent, within the errors. From the CO(2--1) data analysis presented in Section \ref{subsec:mh2},  we find a molecular gas mass of $\log(M_{\text{H}_2}/M_\sun)_N = 11.06^{+0.20}_{-0.39}$ for North and an upper limit of $\log(M_{\text{H}_2}/M_\sun)_S \leq 10.74$ for South. Therefore, we find the two galaxies to be massive star-forming galaxies, with a significant amount of molecular gas and dust. Looking at the mass ratios, we find $\mu= 2^{+3}_{-2}$ for the ratio based on the stellar masses only, and $\mu= 2^{+3}_{-1}$ for the ratio based on the combination of the stellar and gas masses, meaning this system of two galaxies is a major merger (e.g., \citealt{Mantha2018}, \citealt{Duncan2019}).

\begin{table*}
\caption{Properties of the individual components of the dusty major merger.}       
\label{table:properties}      
\centering  
\renewcommand{\arraystretch}{1.5} 
\begin{tabular}{c c c c c c c c c}      
\hline\hline                 
Component & $M_*$ & SFR & $M_{dust}$ & $M_{H2}$ & z & $F_{\mathrm{CO(2-1)}}$ & $F_{\mathrm{CO(5-4)}}$ & $F_{\mathrm{1.1mm}}$\\
 & (1) & (2) & (3) & (4) & (5) & (6) & (7) & (8)\\
\hline                        
   North & $10.95_{-0.05}^{+0.13}$ & $933_{-120}^{+90}$ & $8.82_{-0.07}^{+0.09}$ & $11.06_{-0.39}^{+0.20}$ & $1.17223\pm0.00037$ & $1.56\pm0.25$ & $1.61\pm0.24$ & $1.26\pm0.09$\\
   
   South & $10.59_{-0.08}^{+0.08}$ & $60_{-29}^{+42}$ & $7.56_{-0.33}^{+0.30}$ & $\leq10.74$* & $1.17223\pm0.00037$ & $\leq0.75$* & $\leq0.72$* &$\leq0.28$*\\
\hline                                   
\end{tabular}
\tablefoot{(1) the stellar mass, $M_*$, is in log $M_\odot$; (2) the star formation rate, SFR, is in $M_\odot/\text{yr}$; (3) the dust mass, $M_{dust}$ is in log $M_\odot$; (4) the molecular gas mass, $M_{H2}$ is in log $M_\odot$; (5) the redshift; (6) the CO(2--1) flux in Jy km/s; (7) the CO(5--4) flux in Jy km/s; (8) the dust-continuum in mJy. * indicates 3 $\sigma$ upper limits. All the baryonic properties are derived from SED fitting (see Section \ref{subsec:SEDfitting}), except for the molecular gas mass, which is derived from the ALMA CO(2--1) analysis (see Section \ref{subsec:coflux}).}
\end{table*}

\section{Discussion}
\label{section:discussion}
\subsection{A rare major merger}
Combined with ancillary data, we use ALMA CO(2--1), CO(5--4) and dust-continuum (rest-frame 520$\mu$m) observations with $\leq1.3\arcsec$ angular resolution to reveal a heavily dust-obscured star-forming galaxy in the process of merging with another galaxy, previously identified with HST/F814W observations as a single galaxy. According to \citet{Duncan2019} for the stellar mass range encompassing the galaxies in the system presented in this paper, $\log_{10}(M_*/M_\odot) > 10.3$, $\lesssim 10\%$ of galaxies at $z\sim1$ are in a major merger. This implies that systems like Matilda should be very rare. However, several recent works have shown the limitations of merger visual identification, e.g., \citet{Blumenthal2020} show that more than 50\% of mergers are missed with visual identification of mock Sloan Digital Sky Survey (\citealt{York2000}) g-band images. Furthermore, most of the work on galaxy mergers, e.g., \citet{Tasca2017}, \citet{Ventou2017}, \citet{Duncan2019}, is based on the visual identification of the merger components from rest-frame optical data, making it physically impossible to account for dusty systems such as the one presented here, as dust can partially or totally obscure rest-frame optical data. Although optically faint or dark \textit{single} galaxies are known at high redshifts ($z>3$), as shown in e.g., \citet{Umehata2020}, \citet{Barrufet2023}, \citet{Gomez2023}, the situation for partly optically faint/dark \textit{systems} of galaxies at lower redshifts is unclear. Few works have reported objects that could be similar to the system we study here. One notable example is the work of \citealt{Kokorev2023} that shows a massive 
dusty 
star-forming galaxy at $z = 1.3844$, not fully detected in HST/F814W observations with a potential optically-dark companion. The hypothesis of an optically-dark companion is supported by the tentative 2 $\sigma$ ALMA observations showing a dusty structure reaching towards a secondary star-forming region and a similar CO(2-1) emission line profile as the one we observe in this work. The James Webb Space Telescope (JWST) is also a promising telescope to help us accurately assess the fraction of all galaxy mergers, including dusty systems. In fact, already recent work with JWST data is showing the potential of more systems belonging to this specific class of galaxy mergers. For instance, \citealt{Gillman2023} used JWST/NIRCam observations and identified 5 submillimeter galaxies that could be mergers, with NIR counterparts but no HST optical (HST/F160W) counterparts.

\subsection{The dusty galaxy merger in context of typical star-forming galaxies}
\label{subsec:context}

\begin{figure*}
\centering
    \includegraphics[width=\textwidth]{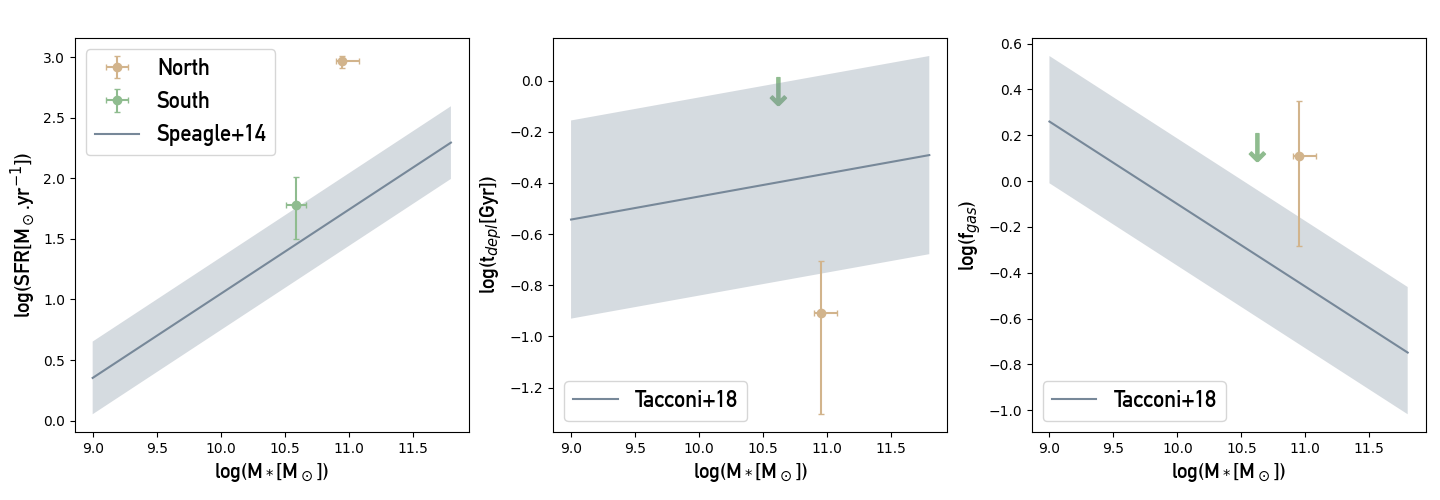}
        \caption{The star formation rate (left), depletion time ($t_{depl} = M_{H_2}/$SFR, centre) and molecular gas fraction ($f_{gas} =  M_{H2}/M_*$, right) of the two galaxies belonging to the dust-obscured merging system compared to scaling relations for these properties of $z=1.17$ main sequence galaxies. North is represented by a beige marker, South is represented by a green marker. With the grey line and shaded area, we show the literature relations for the star-forming main sequence from \citet{Speagle2014} (left) and the depletion time and gas fraction from \citet{Tacconi2018} (centre and right).}
        \label{fig:tacconi}
\end{figure*}

We compare the two galaxy members of the dusty $z=1.17$ galaxy merger to star-forming main sequence galaxies at that same redshift according to \citet{Speagle2014} (Figure \ref{fig:tacconi}). Both galaxies appear above the star-forming main sequence (MS)  (left panel of Figure \ref{fig:tacconi}). North exhibits a particularly high level of star formation, appearing more than 1 dex above the median MS, therefore falling into the regime of starburst galaxies, whereas South is nearly 0.5 dex above that median, consistent with MS galaxies. While the elevated SFR of South can put it in the regime of starburst galaxies, its CO(5--4) to CO(2--1) line ratio seems rather consistent with MS galaxies. To our knowledge, there are no other mergers at $z\sim1$ for which multiple-CO transitions are available for the each member of the merger. Such information is available for isolated galaxies (e.g., \cite{Valentino2020}, \cite{Boogaard2020}, \cite{Harrington2021}), but these typically show more elevated CO line ratios for galaxies with similar SFR-properties as North.\par

We also evaluate the depletion time, i.e.,  $t_{depl} = M_{H2}/$SFR, of the two galaxies according to \citet{Tacconi2018} and compare them to MS $z=1.17$ galaxies (central panel of Figure \ref{fig:tacconi}). South is consistent with MS galaxies, i.e., it shows a star-forming depletion time consistent with the scatter of MS galaxies. We caution that the molecular gas mass derived for South is based on an upper limit measurement of the CO(2-1) flux, therefore South could be forming stars even faster. North is located $\sim 0.5$ dex below the median depletion time relation. Therefore, North is much more efficient at forming stars than South (difference of $\sim 0.8$ dex) and typical $z=1.17$ MS galaxies, by forming stars 3 times faster than the MS galaxies.\par

We determine the molecular gas fraction, i.e., $f_{gas} =  M_{H2}/M_*$, of the two galaxies. The molecular gas fraction indicates how much molecular gas is available to form stars compared to the amount of stars already formed. We again compare this quantity according to the relation for MS $z=1.17$ galaxies from \citet{Tacconi2018} (right panel of Figure \ref{fig:tacconi}). Both North and South appear to exhibit higher molecular gas fractions than MS galaxies, making them rather molecular gas-rich galaxies. Again, we caution that the molecular gas mass derived for South is based on an upper limit measurement of the CO(2-1) flux, therefore the corresponding molecular gas mass fraction is also an upper limit.

The two galaxies part of the dusty galaxy merger are both forming stars with high efficiency, resulting in both galaxies showing above the MS. Although the two galaxies show similar gas fractions, only North, is forming stars rapidly (twice as fast as MS galaxies), resulting in a starburst-like SFR. It is unclear why only one galaxy appears to have the high sSFR of a starburst despite both exhibiting consistently high gas fractions. Possible scenarios involve the orientation and relative rotation directions as shown in simulations (e.g., \citealt{DiMatteo2007}, \citealt{Cox2008}), an AGN boosting the SFR of North (e.g., \citealt{Hopkins2008}, \citealt{Cicone2014}) or simply a difference in the baryonic properties of the two galaxies prior to the merger.

\subsection{The impact of the merger on the properties of galaxies}

Both galaxies in this dusty merger show relatively high levels of star formation and star formation efficiency (see Figure \ref{fig:tacconi}, left and middle panels). Moreover, as mentioned in Section \ref{subsec:mh2}, assuming a lower ${\alpha_\text{CO}}$ value would result in lower molecular gas masses, strengthening even more the star formation efficiencies. Therefore, the star-forming properties of the two galaxies appear consistent with a scenario in which mergers tend to enhance star formation (e.g., \citealt{Kim2009}, \citealt{Saitoh2009}, \citealt{Kaviraj2015}, \citealt{Tacchella2016}, \citealt{Pearson2019}).
However, the merging process could impact each galaxy differently, with North having a significantly higher star formation rate (factor 10 difference), a slightly higher dust mass, and likely a more elevated CO(5-4) to CO(2-1) ratio. The origin of this different impact could also be due to a difference in the evolution and composition of the galaxies prior to their collision.
Similarly, only North exhibits bright dust emission, making it almost entirely invisible in the rest-frame optical wavelengths (see Figure \ref{fig:almaoverhst}). This large amount of dust emission, elevated SFR and high dust and stellar masses could point towards a merger-driven SMG phase (e.g., \citealt{Blain2002}, \citealt{Tacconi2008}, \citealt{McAlpine2019}). However, it remains unclear whether the high brightness of the dust emission is intrinsic to the galaxy or due to the high level of star formation resulting in high heating of the dust.\par

Thanks to the multiple-transition CO observations and their spectral resolution, we find hints of complex dynamics at play in the merger (\ref{subsec:dynamics}). The slight offset in the negative and positive velocities peaks, both in CO(2--1) and CO(5--4), at the location of North (see right column of Figure \ref{fig:PeaksMom0}), suggests signs of rotation of the galaxy, with the northern part moving away from us (negative velocities) and the southern part coming towards us (positive velocities). As for South, it seems to be only traced by the negative velocities peak, both in CO(2--1) and CO(5--4). This suggests that South is moving towards us. The lack of observed rotation as seen for North, could hint to South being face-on, which is consistent with what we observe in the \textit{HST}/F814W image (see Figure \ref{fig:almaoverhst}).

\section{Summary and conclusions}
\label{subsec:summary}
Using ALMA archival data, we have uncovered a dusty galaxy major merger at $z\sim1$ and studied the baryonic properties of the individual galaxies part of this system.

\begin{itemize}
    \item the ALMA CO(2--1), CO(5--4) and dust-continuum observations reveals a dust-obscured galaxy at $z=1.17$;
    \item these observations, complemented by other Subaru, Ultravista, HST, and Spitzer observations, show that the dust-obscured galaxy is merging with another galaxy previously classified (\citealt{Weaver2022}) as a single star-forming galaxy at $z=1.17$;
    \item the two galaxies have a high SFR and a short gas depletion time with respect to literature relations for main-sequence $z=1.17$ galaxies, consistent with a picture in which mergers enhance star formation;
\end{itemize}

With the work presented in this paper, we highlight the necessity for multi-wavelength observations with observatories such as ALMA or JWST to properly assess the fraction of galaxy mergers in our Universe. With their longer wavelengths, ALMA and JWST can open the window onto systems that are otherwise obscured at rest-frame optical and NIR wavelengths. For instance, \citet{Jones2023} use high spatial and spectral resolution with JWST to reveal 4 galaxies, initially disguised as one single massive starburst galaxy \citep{Riechers2013}, showcasing the importance of high resolution to properly account for mergers. The ALMA archive represents the ideal database to find a hidden fraction of galaxy mergers, because the (sub)millimeter wavelengths can observe the dust and the large amount of data increases the probability of finding more systems like the one presented in this paper. JWST, specifically the MIRI instrument offers a high sensitivity, large field of view compared to ALMA ($73.5\arcsec\times112.6\arcsec$) and high spatial resolution, ideal to disentangle faint dusty systems. Due to the different wavelength range probed by ALMA and JWST/MIRI, the observations would trace different types of dust (cold and hot). Therefore, using the two telescopes in synergy would enable a more complete view of the dust in this class of dusty galaxy mergers. With the available angular resolution of the archival ALMA data and our assumptions presented in this work, we were able to use dust-continuum and multiple-CO transitions observations to study of the gas and dust content of the two galaxies involved in this merger serendipitously discovered. Higher resolution observations would enable better informed assumption and even more detailed study of the gas within the galaxies, i.e., the ISM, and its structure and dynamics. 
The gravitational forces between galaxies in a merger can cause significant disruptions to their structures. Tidal forces arise distorting the shapes of galaxies and disrupting their gas, dust, and stars, leaving tidal streams as imprints of the merger process (e.g., \citealt{Stewart2011}, \citealt{Guo2016}, \citealt{Ginolfi2020}). The merger presented in this paper would be an ideal laboratory to study such tidal streams due to its large dust content and proximity  ($\leq$10kpc apart) of the two galaxies.



\begin{acknowledgements}
      We thank the anonymous referee for their careful and constructive report that greatly improved the clarity and quality of this paper. This paper makes use of the following ALMA data: ADS/JAO.ALMA\#2015.1.00260.S  and ADS/JAO.ALMA\#016.1.00171.S. ALMA is a partnership of ESO (representing its member states), NSF (USA) and NINS (Japan), together with NRC (Canada), MOST and ASIAA (Taiwan), and KASI (Republic of Korea), in cooperation with the Republic of Chile. The Joint ALMA Observatory is operated by ESO, AUI/NRAO and NAOJ. This research has made use of the NASA/IPAC Infrared Science Archive, which is funded by the National Aeronautics and Space Administration and operated by the California Institute of Technology. FG acknowledges the support from grant PRIN MIUR 2017-20173ML3WW\_001. "Opening the ALMA window on the cosmic evolution of gas, stars, and supermassive black holes". We thank Nicolas Bouché and Mark Swinbank for useful discussions and feedback.
\end{acknowledgements}

%
%
\bibliographystyle{aa} 
\bibliography{biblio.bib} 






   
  




\begin{appendix} 
\onecolumn
\section{Example of the PhoEBO code}
\label{section:apd}
\begin{figure*}[h]
    \includegraphics[width=\textwidth]{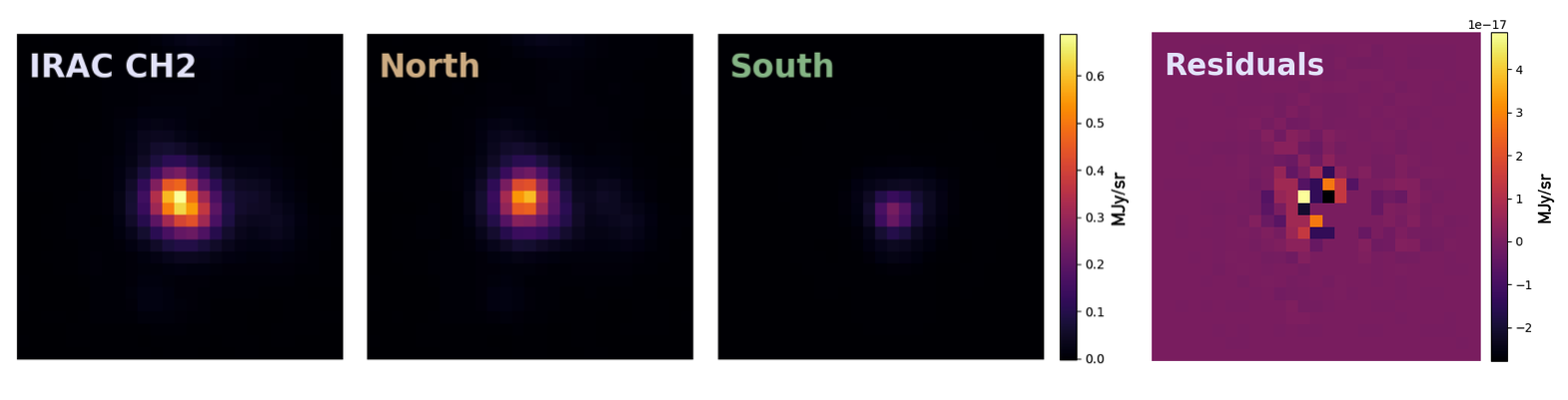}
        \caption{The IRAC/Spitzer Channel 2 maps resulting from the deblending performed by PhoEBO. From left to right: Input map to deblend, output deblended map of North, output deblended map of South, and residuals of the modelling (i.e., input map subtracted from the sum of the two outputs).}
        \label{cl12301}
\end{figure*}
\end{appendix}
\end{document}